\documentstyle[12pt, psfig, leqno]{article}
    
    \def\bc{\begin{center}}
    \def\d{\displaystyle}
    \def\nn{\nonumber}
    \def\e{\varepsilon}
     \def\ec{\end{center}}
     
     \def\p{\partial}
     \def \Rnos {{\rm I\!\rm R}}
     \newtheorem{lem}{Lemma}
    
\begin{document}
\begin{titlepage}
     \bc{\bf LOCALISED SOLUTIONS OF THE DIRAC-MAXWELL EQUATIONS}\ec
     \bigskip

     \bc{\bf C.J. Radford\\
     Department of Mathematics, Statistics and Computing Science\\
     The University of New England\\
     Armidale, N.S.W, 2351\\
     Australia}\\
     chris@neumann.une.edu.au\ec

     \bc{\bf Abstract}\ec

The full classical Dirac-Maxwell equations
 are considered in a somewhat novel form and under various simplifying
assumptions.
A reduction of the equations is performed in the case when the Dirac field is
{\em static}. A further reduction of the equations is made under the assumption
of {\em spherical symmetry}. These static spherically symmetric equations
are examined in some detail and a numerical solution presented.
Some surprising results emerge from this investigation:
\begin{itemize}
  \item Spherical symmetry necessitates the existence of a magnetic monopole.
  \item There exists a uniquely defined solution, determined only by the
demand that the solution be analytic at infinity.
  \item The equations describe highly compact objects with an inner onion like
shell structure.
\end{itemize}
\end{titlepage}

 \bc{\bf INTRODUCTION}\ec

It is an interesting exercise to compare the current development of a quantum
theory of gravitation, from the fully non-linear Einstein theory, to the
development of QED from the linearized Dirac-Maxwell theory. The most startling
difference is the large body of work on the classical, non-linear, theory of
gravitation (General Relativity) -- a theory which includes, in a self
consistent manner, the interactions of the gravitational field itself. There is
no
 comparable body of work on the full Dirac-Maxwell theory (Dirac equations with
electromagnetic interaction, Maxwell equations with Dirac field source - the
so-called ``self interaction''). Of course this situation arose, historically,
because of the rapid development and stunning success of QED.

     Einstein's equations provide a much better description of gravity than
do the linear spin-2 equations. Indeed, one can ``derive'' classical general
relativity from the linear, massless spin-2 theory by summing all the Feynman
diagrams to tree level -- see \cite{du:anomaly} and \cite{de:quant grav}. The
full
Dirac-Maxwell
equations should provide a much better description of electronic matter than
their linearized counterparts (in which self terms are ignored).

Solutions to the Dirac-Maxwell equations are rare, see \cite{ab:book}. There
are
, however, a number of solutions to the Yang-Mills-Dirac
and Yang-Mills-Dirac-Higgs equations,
 see \cite{ma:gaug} and the comprehensive list of references
contained therein.

     The paper is organized as follows:  in \S 1 we write the equations in
      2-spinor form, this description then allows us to (covariantly) solve
the Dirac equations for the electromagnetic potential and so write down a
complete set of equations in terms of the Dirac field.  In \S 2 we examine
the static and spherically symmetric reductions of the equations. In \S 3 we
deal with some general properties of the static spherically symmetric system
and
 in \S 4 present a numerical solution to this system.
     \medskip

     \noindent{\bf \S 1 \ The Dirac-Maxwell Equations}

     In standard notation the Dirac-Maxwell equations are
    \begin{eqnarray}
 \nn &&  \gamma^\alpha (\partial_\alpha -i\,e\,A_\alpha )\psi +
im\psi =0\\
 \nn && F_{\alpha \beta} = A_{\beta ,\alpha} - A_{\alpha , \beta}\\
     &&\p ^\alpha F_{\alpha \beta} = -4\pi e\,j_\beta = -4\pi
e\,\overline{\psi}
\gamma_\beta \psi
    \end{eqnarray}
     Employing the $\gamma_5$-diagonal or van der Waerden  description, see for
example \cite{rp:spin}, we have
     \[ \gamma^{\alpha}=\sqrt{2}\left(\begin{array}{ll}
     0&\sigma^{\alpha}_{B\dot{B}}\\
     \sigma^{\alpha A\dot{A}}&0
     \end{array}\right)\]
with $\sigma^\alpha_{A\dot{A}}$ the van der Waerden symbols,  i.e.
     \begin{eqnarray*}
     (\sigma^0_{A\dot{A}}) &=& \frac{1}{\sqrt{2}}\left(\begin{array}{ll}
     1&0\\
     0&1
     \end{array}\right) \ \ \mbox{and}\\
     (\sigma^j_{A\dot{A}}) &=& \frac{1}{\sqrt{2}} \times\, \mbox{Pauli
Matrix}, \ j=1,2,3.
     \end{eqnarray*}
Where $A, B = 0, 1$ and $\dot{A},\dot{B} = \dot{0}, \dot{1}$ are two-spinor
indices (see [Pe]).

\noindent The Dirac bispinor, $\psi$ is
     \begin{eqnarray*}
 \psi &=& \left(\begin{array}{c}
     u_A\\
     \overline{v}^{\dot{B}}
     \end{array}
     \right)  \ \ \mbox{and}\\ \\
   \overline{\psi} &=& \left(
     v^B , \overline{u}_{\dot{A}}\right).
      \end{eqnarray*}
So that the Dirac equations become
\begin{eqnarray}
\nn &&(\p^{A\dot{A}}-
i\,e\,A^{A\dot{A}})u_A+\frac{im}{\sqrt{2}}\overline{v}^{\dot{A}}=0\\
    &&(\p^{A\dot{A}} + i\,e\,A^{A\dot{A}})v_A + \frac{im}{\sqrt{2}}
\overline{u}^{\dot{A}} =0
\end{eqnarray}
     where $\p^{A\dot{A}} \equiv \sigma^{\alpha A\dot{A}}\p_\alpha$, \
$A^{A\dot{A}}=\sigma^{\alpha A\dot{A}}A_\alpha$.

     The Maxwell equations are
\begin{eqnarray}
\p ^{\alpha}F_{\alpha \beta}
=-4\pi e\,j_\beta =
-4\pi
e\,\sigma_\beta^{A\dot{A}}(u_A\overline{u}_{\dot{A}}+v_A\overline{v}_{\dot{A}})
\end{eqnarray}
  In the linearized theory the ``self current'' $j_\beta$ is ignored.
     We now eliminate the potential $A^{A\dot{A}}$ using (2) -- another
approach is to eliminate ``$A^{\rm self}_\alpha$ ''using the formal Green's
function \cite{ab:spon em} --
 we will use purely algebraic methods.  From the equations (2)
we have
\begin{eqnarray}
\nn \lefteqn{v^A \p^{B\dot{A}}u_B +u^A\p^{B\dot{A}}v_B
+\frac{im}{\sqrt{2}}(v^A\overline{v}^{\dot{A}}+u^A\overline{u}^{\dot{A}})}\\
     && = i\,e  [A^{B\dot{A}}(v^Au_B-u^Av_B)]
\end{eqnarray}

     However, because of the 2-dimensionality of the 2-spinor space we have
     \[ v_Au_B - u_Av_B=\e_{AB}(u^Cv_C) . \]
     Here, $\e_{01} =\e^{01}=1, \ \e_{10}=\e^{10}=-1$,
$\e^{00}=\e_{00}=\e^{11}=\e_{11}=0$; we define $\xi^A=\e^{AB}\xi_B$ and
$\xi_A=\e_{BA}\xi^B$.

     We assume that $u^Cv_C\neq 0$ almost everywhere. Now,
$j^{\alpha}j_{\alpha}=|u^Av_A|^2$, so $u^Cv_C\equiv 0$ implies that the current
vector, $j$, is null -- a massive neutrino field.

We can now solve (4) for the electromagnetic potential $A$,
\begin{eqnarray}
 A^{A\dot{A}}&=&\frac{i}{e(u^cv_c)}\left\{
v^A\p ^{B\dot{A}}
u_B+u^A\partial^{B\dot{A}}v_B+\frac{im}{\sqrt{2}}(u^A\overline{u}^{\dot{A}}+v^A\overline{v}^{\dot{A}})\right\}
\end{eqnarray}
Notice that, from (5), under the gauge transformation
     \[ \left(\begin{array}{l}
     u_A\\
     \overline{v}^{\dot{B}}
     \end{array}\right) \longrightarrow e^{i\chi}\left(\begin{array}{l}
     u_A\\
     \overline{v}^{\dot{B}}
     \end{array}\right) \ \ \ \mbox{we have}  \]
     \[ A_\alpha \longrightarrow A_\alpha + \frac{1}{e}\partial_\alpha \chi
\ \ \ \]
  as we should expect!

     The four complex equations (2) actually over determine the four
\underline{real} quantities $A_\alpha$.  We must impose on (5) the
condition that $A_\alpha$ is real.  These reality conditions can be written
as
  \begin{eqnarray*}
     \overline{(A^{A\dot{A}}u_A\overline{u}_{\dot{A}})} &=&
A^{A\dot{A}}u_A\overline{u}_{\dot{A}}\\
     \overline{(A^{A\dot{A}}v_A\overline{v}_{\dot{A}})} &=&
A^{A\dot{A}}v_A\overline{v}_{\dot{A}}\\
     \overline{(A^{A\dot{A}}u_A\overline{v}_{\dot{A}})} &=&
A^{A\dot{A}}v_A\overline{u}_{\dot{A}}
     \end{eqnarray*}
     With the use of (5) these reality conditions become
\begin{eqnarray}
\nn &&\partial^{A\dot{A}}(u_A\overline{u}_{\dot{A}}) = -
\frac{im}{\sqrt{2}}(u^C v_C -\overline{u}^{\dot{C}}\overline{v}_{\dot{C}})\\
\nn && \partial^{A\dot{A}}(v_A\overline{v}_A)= \frac{im}{\sqrt{2}} (u^C v_C -
\overline{u}^{\dot{C}}\overline{v}_{\dot{C}})\\
 &&u_A\partial^{A\dot{A}}\overline{v}_{\dot{A}} -
\overline{v}_{\dot{A}}\partial^{A\dot{A}} u_A=0
     \end{eqnarray}
     These equations constitute four real first order equations for the four
complex quantities $u^A$ and $v^A$.  A further four real third order equations
for these quantities is obtained upon substitution of (5) into the Maxwell
equations (3).  Note that adding the first two equations of (6) leads
to the equation of conservation for $j^\alpha$.

     \bigskip

     \noindent{\bf \S2 \ Reduction of the System}

     \noindent\underline{2.1 The Static Equations}

     Firstly, we impose the condition that the field is static.  We assume
that there exists a Cartesian Lorentz frame in which $j^\alpha =
\delta^\alpha_0j^0$.  Imposing this condition one quickly finds that
     \[ v^A=e^{i\chi}
\sqrt{2}\sigma^{0 A\dot{A}}\overline{u}_{\dot{A}},\ \mbox{with $\chi$ a real
function.} \]
     the current vector is now
     \[ j^\alpha = \sqrt{2}(u^0\overline{u}^{\dot{0}}+u^1
\overline{u}^{\dot{1}})\delta^\alpha_0 \]
     Now write
     \[ u^A =e^{-imt}\zeta^A .\]
     The reality conditions are
\begin{eqnarray*}
&&\partial_{A\dot{A}}(\zeta^A\overline{\zeta}^{\dot{A}}) =
\frac{-2m}{\sqrt{2}}(|\zeta^0|^2 +|\zeta^1|^2)\sin \chi\\
&&(\partial_{0\dot{0}}+\partial_{1\dot{1}})(|\zeta^0|^2+|\zeta^1|^2)=0\\
&&   \zeta^0(\partial_{0\dot{0}}+\partial_{1\dot{1}})\zeta^1-
\zeta^1(\partial_{0\dot{0}}+\partial_{1\dot{1}})\zeta^0 =
     i[\zeta^0\zeta^1(\partial_{0\dot{0}}-
\partial_{1\dot{1}})+(\zeta^1)^2\partial_{1\dot{0}}-
(\zeta^0)^2\partial_{1\dot{1}}]\chi
   \end{eqnarray*}
     the expressions for the potential $A^{A\dot{A}}$ can now be written
down, although we won't do this at this stage.

     Now under a gauge transformation  we have
     \begin{eqnarray*}
     \zeta^A &\to & e^{i\mu} \zeta^A \ \ {\rm and}\\
     A_\alpha &\to & A_\alpha +\frac 1e\partial_\alpha \mu.
     \end{eqnarray*}
     We fix the gauge by defining real functions $X,Y$ and $\eta$ as follows
     \begin{eqnarray*}
     \zeta^0  &=& X\,e^{\frac i2 (\chi + \eta )}\\
     \zeta^1&=& Y\,e^{\frac i2 (\chi -\eta )}
     \end{eqnarray*}

    Our equations can be given in a particularly suggestive {\em three vector}
 form by writing (in our Cartesian coordinates)
     \begin{eqnarray*}
      \ell & = & (\sigma^{\alpha}_{ A \dot{A}} u^A \overline{u}^{\dot{A}})\\
             & = & (\ell^0,\frac{1}{\sqrt{2}}{\mbox {\boldmath $V$}})
     \end{eqnarray*}
with
     \[\ell^0 = \frac{1}{\sqrt{2}} (X^2+Y^2)\] and
     \[{\mbox {\boldmath $V$}} = (2 XY\cos \eta, \ 2XY\sin \eta, X^2-Y^2) \]
The reality conditions become
\begin{eqnarray*}
&&\frac{\partial}{\partial t} (X^2+Y^2)=0\\
&&     {\mbox {\boldmath $\nabla . V$}} = -2m(X^2+Y^2)\sin \chi\\
&&     \frac{\p {\mbox {\boldmath $V$}}}{\partial t} + ({\mbox {\boldmath
$\nabla$}}\chi )
{\mbox {\boldmath $\times  V$}}= {\mbox {\boldmath $0$}}
     \end{eqnarray*}
With electromagnetic potential
    \begin{eqnarray*}
   &&A^0 =\frac me(\cos \chi -1)+\frac{(X^2-
Y^2)}{2\,e(X^2+Y^2)}\frac{\partial\eta }{\partial
t}+\frac{({\mbox {\boldmath $\nabla$}} \chi){\mbox {\boldmath $. V$}}}{2e(X^2 +
Y^2)}\\
     &&{\mbox {\boldmath $A$}} = \frac{1}{2e(X^2 + Y^2)} \left[ \frac{\partial
\chi}{\partial t} {\mbox {\boldmath $V$}} + (X^2 - Y^2) {\mbox {\boldmath
$\nabla$}} \eta -
{\mbox {\boldmath $\nabla \times V$}} \right]\\
     &&{\rm where} \ \ {\mbox {\boldmath $A$}} = (A^1, A^2, A^3)
     \end{eqnarray*}
The full system is given by the above two sets of equations and the Maxwell
equations.

\noindent\underline{2.2 Spherical Symmetry}

     We now impose spherical symmetry upon our static system.  A minimal
requirement that the Dirac field be static and spherically symmetric (in any
gauge) is that the vector $\ell$, above, is spherically symmetric. We require
    \[ \left[ X_i, \ell \right] = 0\]
     where the $X_i, i = 1, 2,3$, are the three (vector) generators of
rotations.

     These conditions imply that $\ell$ has time and radial components only
and that these components are functions of $(t,r)$ only, $r = \sqrt{(x^1)^2 +
(x^2)^2 +
(x^3)^2}$.

     Using the notation above we have
     \begin{eqnarray*}
     X^2 + Y^2 & = & R = R(r) \ {\rm , only}\\
     {\mbox {\boldmath $V$}} & = & {\mbox {\boldmath $|V|$}} \hat{{\mbox
{\boldmath $r$}}} = R \hat{{\mbox {\boldmath $r$}}}
     \end{eqnarray*}
     where $\hat{{\bf r}} = (\sin \theta \cos \phi, \sin \theta \sin\phi,
\cos \theta)$ in terms of the polar coordinates $r, \theta, \phi$.
     We have
     \begin{eqnarray*}
     X & = & \sqrt{R} \cos(\theta/2)\\
     Y & = & \sqrt{R} \sin (\theta/2)\\
     \eta & = & \phi
     \end{eqnarray*}
The Dirac bispinor is now
\[\psi = e^{-imt}\,\sqrt{R}\left( \begin{array}{c}
          -e^{\frac{i}{2}(\chi -\phi)} \sin (\frac{\theta}{2})\\
           e^{\frac{i}{2}(\chi +\phi)} \cos (\frac{\theta}{2})\\
          -e^{\frac{-i}{2}(\chi +\phi)} \sin (\frac{\theta}{2})\\
           e^{\frac{-i}{2}(\chi -\phi)} \cos (\frac{\theta}{2})
                \end{array}\right).\]

     The equations are now as follows

   \begin{eqnarray}
  \nn &&\chi = \chi(r), \ R=R(r)\\
   \nn && \mbox{{\boldmath $A$}}=\frac{1}{2\,e}
      \frac{\cot \theta}{r}\hat{{\mbox {\boldmath $\phi$}}}\\
   \nn && A^0=\frac me (\cos \chi -1) + \frac{1}{2\,e}\frac{d\chi}{dr}\\
  \nn &&\frac{d}{dr}(r^2R) =-2mr^2R\sin \chi\\
     &&\frac{d}{dr}(r^2\frac{dA_0}{dr})=-4\sqrt{2}\,\pi er^2R
     \end{eqnarray}
     The really surprising result here is the unavoidable appearance of the
magnetic monopole term
     \[ {\mbox {\boldmath $A$}} =\frac{1}{2\,e}\frac{{\em
cot}\,\theta}{r}\hat{{\mbox {\boldmath $\phi$}}}. \]
Here $\hat{{\mbox {\boldmath $\phi$}}}$ is the usual azimuthal unit vector, in
terms of a coordinate basis $ {\mbox {\boldmath $A$}}
=-\frac{1}{2 e}\cos \theta \,d \phi$.

  We should also impose a normalisation condition (or finite total charge
condition) on any solution
     \[ \int j^\alpha dS_\alpha < \infty \]
     on any space like hypersurface.  This leads to the condition
     \[ \int^\infty_{r=0} r^2R\,dr < \infty \]

     To end this section we rewrite the determining radial equations in a
more transparent form by introducing the following new (dimensionless)
variables
     \begin{eqnarray*}
     \rho &=& 2mr\\
     a&=& \frac em A^0\\
     q&=& 4\pi \sqrt{2}\frac{e^2}{m}r^2R.
     \end{eqnarray*}

    \begin{eqnarray}
  \nn && \frac{d\chi}{d\rho} =a+1-\cos \chi\\
   \nn &&\frac{d}{d\rho} (\rho^2\frac{da}{d\rho}) =-q\\
   &&\frac{dq}{d\rho} =-q\sin \chi
     \end{eqnarray}

     \medskip

     \noindent{\bf \S 3 \ Static Spherical Symmetry}

     \noindent\underline{3.1 Some General Properties}

The system of equations (8) possesses the discrete symmetry
\begin{eqnarray}
\nn &&\chi \to \pi - \chi \\
\nn &&a+1 \to -(a+1) \\
&&q \to -q
\end{eqnarray}
This is just the operation of charge conjugation; $q$ needs to be reinterpreted
(it was originally defined as non-negative) to account for the change in sign
 of the charge which manifests itself on the right hand side of the second
equation of (8). We write $q = \epsilon Q$, with $\epsilon ^2 = 1$ and
$Q \geq 0$, so $Q= \sqrt{2}\frac{e^2}{m}r^2R$; then,
under (9), we have $\epsilon \to -\epsilon$ and $Q \to Q$.

Our equations read,
\begin{eqnarray}
\nn&&\frac{d\chi}{d\rho}=a+1-\cos \chi\\
\nn&&\frac{da}{d\rho}=-\epsilon f/\rho^{2}\\
\nn&&\frac{df}{d\rho}=Q\\
&&\frac{dQ}{d\rho}=-Q\sin \chi
\end{eqnarray}
Where we have introduced the new variable $f$ to give a set of four first
order, ordinary differential equations. This new variable is directly related
to the
magnitude total (Dirac field) charge contained in a ball of radius $r$, $B(r)$,
    \begin{eqnarray*}
  e\int_{B(r)} j^\alpha dS_\alpha &=&e\int_{B(r)}j^{0}d^3x\\
                                   &=&4\pi e\sqrt{2}\int^{r}_{s=0}s^2R(s)\,
ds\\
                                   &=&\frac{2\pi}{e}\int^{\rho}_{\sigma =0} Q\,
d\sigma\\
                                   &=&\frac{1}{2 e}(f(\rho )-f(0))
     \end{eqnarray*}
In view of this -- and the fact that $\frac{df}{d\rho}= Q =\epsilon q$ is
proportional to the charge density on a shell of radius $r$ -- we will use the
following condition on our system (10).

\[\mbox{{\bf (C1)}}\quad\quad\quad \left[ \begin{array}{l}
\mbox{On}\, \rho>0\,, f \, \mbox{is a bounded} \, C^1 \, \mbox{function, with
bounded first derivative.}\\
\d{\mbox{Both}\, f \,\mbox{and}\, \frac{df}{d\rho}\,\mbox{have well defined
limits as}\, \rho \to \infty}
                           \end{array}\right.\]

We will now develop some qualitative results which indicate the types of
solution which can exist under rather general (and physically reasonable)
conditions.

\begin{lem} Suppose $(\chi,a,f,Q)$ is a solution of (10) on $\rho >0$
, then under {\bf C1} the function $a$ is $C^2$ on $\rho>0$ with $a$ and
$\frac{da}{d\rho}$ bounded on intervals $\rho\geq \rho_{1}>0$ and $\rho a$
bounded on
$0\leq \rho \leq \rho_{1} <\infty$. If $f(0)\neq 0$ or
$\frac{df}{d\rho}(0)=Q(0)>0$
then $a$ is unbounded as $\rho \to 0$.
\end{lem}

\noindent {\bf Proof.} We first establish that $Q$ and $f$ have well defined
limits as $\rho$ approaches $0$. We are assuming that the solution
$(\chi ,a,f,Q)$ exists on $\rho>0$, so for $\rho_{2}> \rho_{1} >0$
and using equations (10), we have
\begin{eqnarray*}
 |Q(\rho_{2})-Q(\rho_{1})| &=&|\int^{\rho_{2}}_{\rho_{1}}Q(\sigma) \sin \chi
(\sigma)\,d\sigma |\\
                         &\leq&\int^{\rho_{2}}_{\rho_{1}}|Q \sin \chi |
d\sigma\\
                         &<& M_{1}(\rho_{2}-\rho_{1})\,\mbox{,
here}\,M_{1}=\sup_{\Rnos ^{+}}Q<\infty .
\end{eqnarray*}
Letting $\rho_{1}$ and $\rho_{2}$ approach zero, we have
$|Q(\rho_{2})-Q(\rho_{1})| \to 0$. Using Cauchy's criterion we conclude that
$Q(0^+)$ exists. A similar
 argument can be given to demonstrate the existence of $f(0^+)$.

\noindent The boundedness of $\frac{da}{d\rho}$, on $\rho \geq \rho_{1}$
follows from
\[\epsilon \frac{da}{d\rho} = -f/\rho^2 \]
Integrate this expression to bound $a$ on $\rho \geq \rho_{1}>0$.

\noindent Write $\Omega =\epsilon \rho a$, then
\begin{eqnarray*}
\rho \frac{d\Omega}{d\rho} - \Omega &=& -f \mbox{, and}\\
\rho \frac{d^2\Omega}{d\rho^{2}}&=&-Q
\end{eqnarray*}
{}From the second of these equations we have, on $(0,\rho_1]$,
\[M_{1}\ln (\frac{\rho_{1}}{\rho})+\frac{d\Omega}{d\rho}(\rho_{1})\geq
\frac{d\Omega}{d\rho}(\rho)\geq \frac{d\Omega}{d\rho}(\rho_{1})\]
So we have (note as $a$ is $C^2$ away from $\rho =0$ so is $\Omega$)
\[\rho \frac{d\Omega}{d\rho} \to 0 \mbox{, as}\, \rho \to 0\]
Hence from the first of the $\Omega$ equations we see that $\Omega$ has a
well defined limit as $\rho \to 0$, in fact
         \[ \lim_{\rho \to 0} \Omega (\rho) = f(0)\]
An immediate consequence is that if $f(0) \neq 0$ then $a$ is unbounded as
$\rho \to 0$.

\noindent Now from (10) we have $Q \geq Q(0)e^{-\rho}$, so
that $f(\rho)-f(0)\geq \frac{df}{d\rho}(0)(1-e^{-\rho})$ -- recall
$\frac{df}{d\rho}=\epsilon q = Q$.
Given the earlier result we may take $f(0)=0$ -- otherwise $a$ is unbounded.We
can now integrate our $\frac{da}{d\rho}$ equation with this bound for $f$,
\[\epsilon (a(\rho)-a(\rho_1))\geq \frac{df}{d\rho}(0)\left[
\frac{(1-e^{-\rho})}{\rho}-\frac{(1-e^{-\rho_{1}})}{\rho_{1}}\right]+\frac{df}{d\rho}(0)\int ^{\rho_{1}}_{\sigma =\rho} \frac{e^{-\sigma}}{\sigma}d\sigma\]
As $\rho \to 0$ the integral on the right side of the inequality diverges to
$+\infty$.
\bigskip

\begin{lem} Suppose $(\chi,a,f,Q)$ is a solution of (10) on $\infty>\rho>0$,
then under {\bf C1} the function $\chi$ is $C^1$ with $\frac{d\chi}{d\rho}$
bounded on intervals $\rho\geq \rho_{1}>0$. If $f(0)\neq 0$ then $\chi$ is
unbounded as $\rho \to 0$.
\end{lem}

\noindent {\bf Proof.} The regularity of $\chi$ -- on its presumed interval of
existence -- is established using standard
theory (see, for example, \cite{zh:de} or \cite{st:de}) after first noting that
the right side of
\[ \frac {d\chi}{d\rho} = a+1-\cos \chi \]
is $C^2$ in $\rho$ (treating $a$ as a known function and using lemma 1) and
$C^{\infty}$ in $\chi$.

\noindent From the above equation we also have
\[\epsilon (a+1)-1 \leq \epsilon \frac{d\chi}{d\rho} \leq \epsilon (a+1)+1 \]
which gives the required bounds (using lemma 1).
\noindent Working on $(0,\rho_{1})$ we have (as in the proof of lemma 1)
\[C_{0}+\frac{f(0)}{\rho}\leq \epsilon \frac {d\chi}{d\rho} \leq
C_{1}+\frac{f(\rho_{1})}{\rho} \]
where $C_{0} = \epsilon (a(\rho_{1})+1)-1-\frac{f(0)}{\rho_{1}}$ and
$C_{1} = \epsilon (a(\rho_{1})+1)+1-\frac{f(\rho_{1})}{\rho_{1}}$.

\noindent Integrating,
\[ C_{2}-C_{1}(\rho_{1}-\rho)+f(\rho_{1})\ln (\rho) \leq \epsilon \chi \leq
C_{3}-C_{0}(\rho_{1}-\rho)+f(0)\ln (\rho) \]
with $C_{2}= \epsilon \chi (\rho_{1})-f(\rho_{1})\ln (\rho_{1})$ and $C_{3}=
\epsilon \chi(\rho_{1})-f(0)\ln (\rho_{1})$.
If $f(0)<0$ choose $\rho_{1}$ near $0$ so that $f(\rho_{1})<0$ and we have from
our last inequality that $\epsilon \chi \to \infty$ as $\rho \to 0$. If
$f(0)>0$
our inequality yields $\epsilon \chi \to -\infty$ as $\rho \to 0$.
\bigskip

There is one other condition which makes sense ``physically'': if we have an
isolated system we expect the charge density should go to zero at infinity.
\[\d{\mbox{{\bf (C2)}}\quad\quad\quad \left[ \begin{array}{l}
 \d{\frac{df}{d\rho}=Q \to 0,\, \mbox{as}\, \rho \to \infty.}\\
           \end{array}\right.}\]
\begin{lem} Suppose $(\chi,a,f,Q)$ is a solution of (10) on $\rho>0$ under
conditions {\bf C1} and {\bf C2}. Then $-2 \leq a_{\infty} \leq 0$
 where $a \to a_{\infty}$ as $\rho \to \infty$.
\end{lem}

\noindent {\bf Proof.} We first establish that $a$ has a well defined limit,
written as $a_{\infty}$, as $\rho \to \infty$; for $\rho_{2}>\rho_{1}>0$,
we have
\begin{eqnarray*}
|a(\rho_{2})-a(\rho_{1})|&=&|\int^{\rho_{2}}_{\rho_{1}}\frac{f(\sigma)d\sigma}{\sigma^2}|\\
&<&M_{2}(\frac{1}{\rho_{1}}-\frac{1}{\rho_{2}})\,\mbox{, where}\,
M_{2}=\sup_{\Rnos^+}|f|.\\
\end{eqnarray*}
Letting $\rho_{1}$ and $\rho_{2}$ approach $\infty$, we conclude that the limit
$a \to a_{\infty}$ exists.

\noindent From the first and second equations of (10) we have
          \begin{eqnarray*}
            \frac{d}{d\rho}\ln |a+1-\cos \chi| &=&(\frac{da}{d\rho}+\sin \chi
\frac{d\chi}{d\rho})/(a+1-\cos \chi)\\
                                               &=&\sin \chi -\frac{f}{\rho^2
(a+1-\cos \chi)}
          \end{eqnarray*}
Consequently, on $\rho \geq \rho_{1}>0$, integrating the last equation of (10)
we have
           \begin{eqnarray*}
-\ln \left[Q/Q(\rho_{1})\right]&=&\int^{\rho}_{\rho_{1}}\sin
\chi(\sigma)\,d\sigma
\\
                              &=&\ln \left|\frac{a+1-\cos
\chi}{a(\rho_{1})+1-\cos
\chi(\rho_{1})}\right|+\int^{\rho}_{\rho_{1}}\frac{f(\sigma)\,
d\sigma}{[\sigma^2(a(\sigma)+1-\cos \chi(\sigma))]}
          \end{eqnarray*}
Now assume $a_{\infty}>0$. In fact, this also takes care of the case
$a_{\infty}<-2$, since under the discrete conjugation symmetry, (9), $a \to
-(a+2)$ and
in particular $a_{\infty}\to -(a_{\infty}+2)$.

\noindent Working on $\rho \geq \rho_{1}>0$ we have, for $\rho_{1}$ large
enough,
\[a(\rho)>\alpha_{0}>0\,\mbox{, for}\, \rho \geq \rho_{1}\,\mbox{ and some
constant}\, \alpha_{0}.\]
Then, as $a \leq \frac{d\chi}{d\rho}\leq a+2$, so
\[0< \alpha_{0}< \frac{d\chi}{d\rho}<M \, \mbox{, where}\, M \,\mbox{is a
finite constant -- see lemma 1.}\]
\[\mbox{i.e.,}\; \alpha_{0}<a+1-\cos \chi < M\]
Clearly both terms on the right side of our equation for $-\ln \left[
Q/Q(\rho_{1})\right]$ are bounded. This contradicts our assumption {\bf C2}
that $Q \to 0$ as
$\rho \to \infty$.

\bigskip

The constant $a_{\infty}$ can be removed from the potential via a gauge
transformation. Under $\psi \to e^{-ima_{\infty} t}\psi$, we have $a \to
a-a_{\infty}$ .
After this transformation the Dirac field $\psi$ has time dependence
$e^{-iEt}$, where $-m \leq E=(1+a_{\infty})m \leq m$.

The three lemmas give a basic characterisation of the solutions obeying {\bf
C1}
 and {\bf C2}. If $f(0)\neq 0$ or $Q(0)>0$ then $a$ diverges at the origin,
 these are solutions which can be pictured as an Dirac field surrounding a
central charged monopole -- the numerical solution of \S 4 is of this type.
There is also the possibility of solutions with $a$ and $R$ (recall:
$Q=\sqrt{2}\,\frac{e^2}{m}r^2R$) everywhere bounded, such solutions were
suggested by the
work of Wakano \cite{wa:dirac-max}, who examined numerical solutions for what
could be called ``half linearised'' Dirac Maxwell equations
 -- ``half linearised'':
 if the electrostatic potential is ``dominant'' ignore the Maxwell equation
involving the electromagnetic vector potential and {\em vice versa}. In fact,
 as the following theorem demonstrates, no such solutions exist.

\noindent {\bf Theorem}.{\em There does not exist a non-trivial solution of
(10)
on} $\rho \geq 0$ {\em under conditions {\bf C1} and {\bf C2} with} $a$
{\em and} $P= Q/\rho^2$ {\em bounded on} $\rho \geq 0$.

\noindent{\bf Proof.} From lemma 1 we have, $f(0)=Q(0)=0$.
\noindent Next we establish that both $\chi$ and $a$ have well defined limits
as
$\rho \to 0$. Note that $\frac{d\chi}{d\rho} = a+1-\cos \chi$ is bounded, under
the hypothesis of the theorem, as $\rho \to 0$; hence, by an argument of the
sort used previously, $\chi \to \chi (0) =\chi_0$, say, as $\rho \to 0$. We
also
have
\[f=\int^{\rho}_{\sigma=0}Q(\sigma)\, d\sigma =\int^{\rho}_{\sigma=0}\sigma^{2}
P(\sigma)\,d\sigma \geq 0\]
Using the mean value theorem we have, for some $\rho_{1}$, $\rho>\rho_{1}>0$
\[f(\rho)=\rho\rho_{1}^2P(\rho_{1})<\rho^3M_{3}\,\mbox{, where}\, M_{3}=
\sup_{\Rnos^+}P<\infty\]
Thus,
\begin{eqnarray*}
|a(\rho_{2})-a(\rho_{1})|&=&\left|
\int^{\rho_{2}}_{\rho_{1}}\frac{f(\rho)\,d\rho}
{\rho^2}\right|\\
                         &<& \int^{\rho_{2}}_{\rho_{1}}\rho M_{3}\,
d\rho=\frac{1}{2}M_3 (\rho_{2}^2-\rho_{1}^2)
\end{eqnarray*}
Letting $\rho_1$, $\rho_2 \to 0$ establishes the existence of the limit for
$a$,
we write
\[\lim_{\rho \to 0}a=a_{0}\]
Now we use an argument similar to that used in the proof of lemma 3 to show
that
$\frac{d\chi}{d\rho}=a+1-\cos \chi \to 0$, as $\rho \to 0$. On $(0,\rho_1)$ we
have
\[(*)\; -\ln [Q(\rho_1)/Q(\rho)]=\ln \left|\frac{a(\rho_{1}+1-\cos \chi
(\rho_{1})}{a+1-\cos \chi}\right|+\int^{\rho_{1}}_{\sigma
=\rho}\frac{f(\sigma)\,d\sigma}{\sigma^2[a(\sigma)+1-\cos \chi (\sigma)]}\]
Assume, $\frac{d\chi}{d\rho}(0)=a_{0}+1-\cos \chi_{0}\neq 0$. Then, choosing
$\rho_{1}$ near $0$ so that
$\frac{d\chi}{d\rho}(\rho)\frac{d\chi}{d\rho}(0)>0$, for
$\rho_{1}>\rho >0$, we have
\[ \left|\int^{\rho_{1}}_{\rho}\frac{f(\sigma)\, d\sigma}{\sigma^{2}[a(\sigma)
+1-\cos \chi (\sigma)]}\right|<M_{3}\int^{\rho_{1}}_{\rho}\frac{\sigma d\sigma}
{\left|\frac{d\chi}{d\rho}(\sigma)\right|}\]
The right side of this inequality is bounded as $\rho \to 0$. Consequently, the
right side of $(*)$ is bounded as $\rho \to 0$. But this contradicts the
assumption of the theorem that $Q=\rho^{2}P \to 0$ as $\rho \to 0$. Thus
\begin{eqnarray*}
 \frac{d\chi}{d\rho}(0)&=&a_0+1-\cos \chi_0 =0,\,\mbox{or}\\
                    a_0&=&-1+\cos \chi_0
\end{eqnarray*}
We now assume $\epsilon =+1$, the case $\epsilon =-1$ can (of course!) be
obtained by conjugation. As $\frac{da}{d\rho}=-f/\rho^2 <0$ on $\rho>0$, so
\[a_{0}=-1+\cos \chi_{0}>a>a_{\infty} \geq -2\]
on $\rho>0$.

\noindent Define new variables
\begin{eqnarray*}
  U&=& \sqrt{Q} \cos (\chi /2),\, \mbox{and}\\
  V&=& \sqrt{Q} \sin (\chi /2).
\end{eqnarray*}
We have
\begin{eqnarray*}
  \frac{dU}{d\rho}&=&-\frac{1}{2}(a+2)V, \, \mbox{and}\\
  \frac{dV}{d\rho}&=&\frac{1}{2}aU.
\end{eqnarray*}
The pair $U$ and $V$ also satisfy the following linear, second order equations
\begin{eqnarray}
\nn &&\frac{d^2U}{d\rho^{2}}+\frac{f}{\rho^2
(a+2)}\frac{dU}{d\rho}+\frac{1}{4}a
(a+2)U=0,\, \mbox{and}\\
&&\frac{d^2V}{d\rho^2}+\frac{f}{\rho^2 a}\frac{dV}{d\rho}+\frac{1}{4}a(a+2)V
     =0
\end{eqnarray}
We note that $\frac{f}{\rho^2 (a+2)}$is bounded on intervals $[0,\rho_2]$,
with $0\leq \rho_2 <\infty$ and that $\frac{f}{\rho^2 a}$ is bounded on
$(0,\infty]$ -- in the first case we may have $a_{\infty}=-2$,
 whereas we may have $a_0=0$ in the second; we also have
\[\frac{1}{4} a (a+2) \leq 0,\, \mbox{on,}\, \rho \geq 0.\]
{}From the definitions of $U$ and $V$
 we have $U$, $V$ $\to 0$ as $\rho \to 0$ or $\infty$. Thus, by the
maximum principle for odes (see \cite{pr:max}), we conclude $U \equiv V\equiv
0$
, so $Q \equiv 0$.
\noindent There do not exist non-trivial solutions.
\bigskip

  \noindent{\bf \S4 \ Numerical Solutions}

Numerical solutions to the system (10), with $\epsilon =1$, were sought by
first expanding in a power series from either $s=0$
($\rho =\infty$, with $s=1/\rho$)
or $\rho=0$ and then evolving the system in $s$ or $\rho$, respectively, using
a
 M\small ATLAB\normalsize \,interface to the NAG library, \cite{ma:mat-nag}.

   \noindent\underline{4.1 Solutions Near $\rho = 0$}

In lemma 1 we found that $\Omega =\rho a$ was bounded, near $\rho =0$, with
          \[\lim_{\rho \to 0}\Omega (\rho) =f(0)\]
It is natural then to seek solutions of the form $a=\Omega (\rho)/\rho $, near
$\rho =0$, with $\Omega$ analytic in $\rho$. From (10) it can be seen that
 both $f$ and $Q$ must be analytic, with $Q(0)=0$.

However, with $Q$ analytic and $Q(0)=0$, the last equation of (10) can only
be satisfied with $Q \equiv 0$ -- which implies $a=c_0+c_1 /\rho$, we will
refer to such solutions as trivial. The behaviour of the system near $\rho =0$
may be
 quite complex; from the proof of the theorem (see equations (11)), with
$a = \Omega /\rho$, we see that near $\rho =0$ the second order equations
for $U$ and $V$ have indicial equation (see \cite{in:ode})
      \[\lambda^2+\frac{1}{4}\Omega (0)^2 =0\]
This implies that $U$ and $V$ have behaviour
\[ U \mbox{\, or \, }V \sim \omega_1 (\rho) \cos (\frac{\Omega (0)}{2}\ln \rho
)
+\omega_{2} (\rho) \sin (\frac{\Omega (0)}{2} \ln \rho) \mbox{, near\,}\rho
=0\]

   \noindent\underline{4.2 Solutions Near $\rho = \infty$}

Near $\rho =\infty$ we expect $a \sim a_{\infty}+\frac{c_1}{\rho}+
\frac{c_2}{\rho^2}+\cdots $. Assuming $a$ is analytic in $s=1/\rho$,
near $s=0$, then implies that $f$, $Q$ and $\sin \chi$ are also analytic in
$s$,
in fact
   \begin{eqnarray*}
          && f=\frac{d a}{d s} \\
         && Q=-s^2 \frac{d^2 a}{d s^2}\mbox{\, , and}\\
  && \sin \chi=2s+s^2\left( \frac{d^3a/ds^3}{d^2a/ds^2}\right)
   \end{eqnarray*}
With the assumption that $\chi$ is analytic near $s=0$ a {\em uniquely} defined
power series results if we demand that the solution be non-trivial ($\chi$ does
 have the freedom to add integer multiples of $2\pi$).

\noindent The resulting power series has no free parameters, it is uniquely
determined.

\noindent The lower order portion of the
power series solution is as follows
      \begin{eqnarray*}
     \chi &=&2 s+\frac{1}{21}s^3+\frac{3}{520}s^5 +O(s^7)\\
        a &=& -4s^2+\frac{3}{7}s^4-\frac{341}{5096}s^6+O(s^7)\\
        f &=& -8s+\frac{12}{7}s^3-\frac{1023}{2548}s^5+O(s^7)\\
        Q &=& 8s^2-\frac{36}{7}s^4+\frac{5115}{2548}s^6+O(s^7)
      \end{eqnarray*}
\noindent Using the power series to determine initial conditions it was found
that the numerical results were very stable for a good range of initial values
for $s$ ($s_0=.000001$ to $s_0=.01$), the results were somewhat unstable for
$s_0<0.0000001$. The results presented in figures 1 to 4 were obtained by first
shooting from near $s=0$ towards $\rho=0$ and then using the final values of
this run
as initial conditions to shoot from near $\rho=0$ towards $s=0$, to verify the
solution.

In figures 1 to 4 $\chi$, $a$, $f(r)-f(0)$ (proportional to the ``electron''
 charge interior to a ball radius $r$) and $Q$ are plotted against the radial
distance measured in units of the Compton wavelength
(i.e. against $\frac{1}{2}\rho=mr$).

\noindent {\em Interpretation}: The solution represented in figures 1 to 4 can
be thought of as a central, charged monopole (point source), surrounded by an
oppositely charged Dirac field -- near $\infty$ the electrostatic potential
 behaves as $A^0 = \frac{m}{e} a \sim \frac{-1/(m e)}{r^2}$ and near $r=0$ the
potential behaves as $A^0 \sim  \frac{-\alpha /e}{r}$ (where $\alpha \approx
5.7037$ is the
magnitude of the slope of the line in figure 5, where $a$ is plotted against
 $1/(m r)$).

\vspace{5 mm}

\psfig{figure=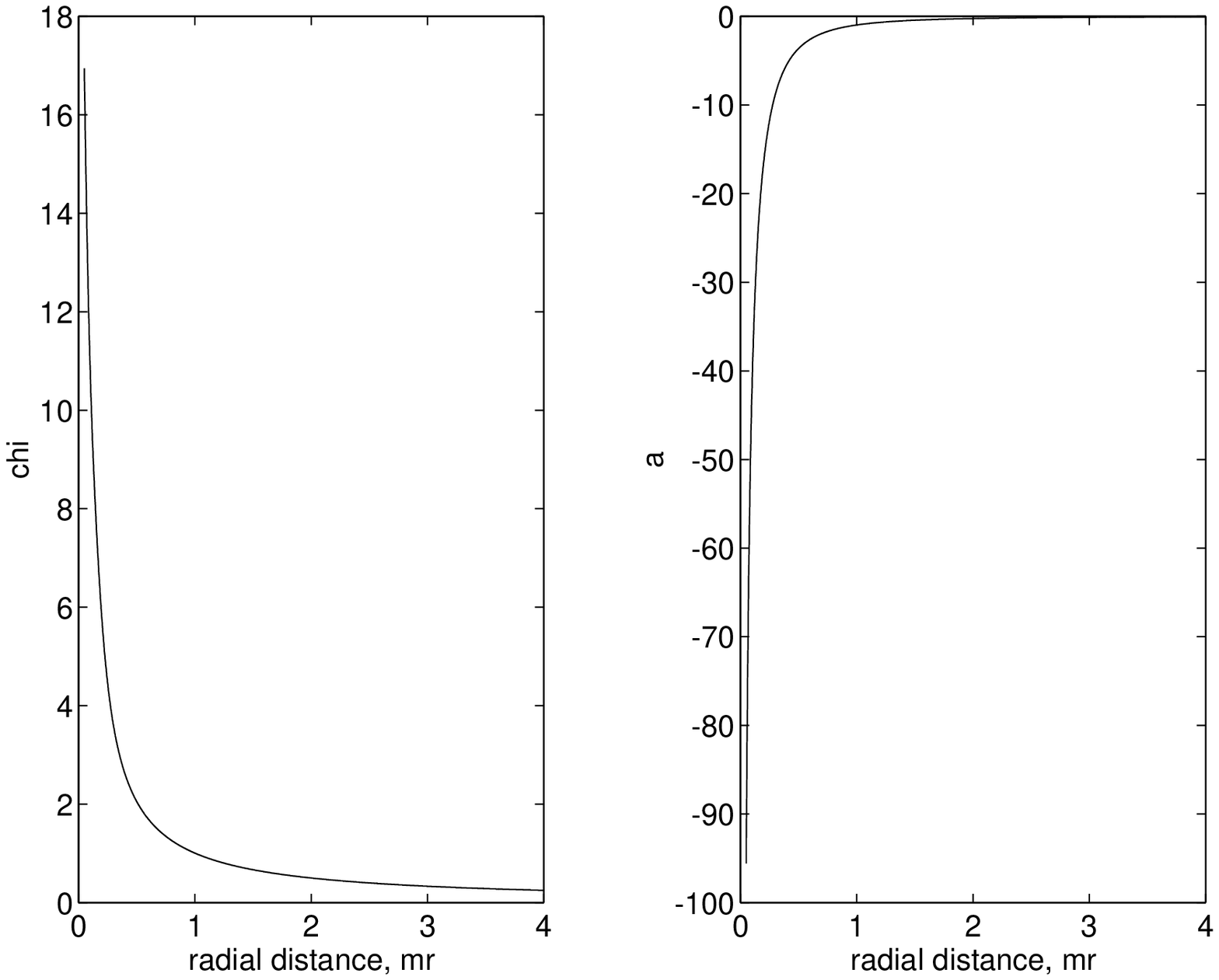,height=7cm,width=15cm}
\begin{tabular}{lll}
Figure 1: The Angular Variable, $\chi$.&\,\hspace{.8 cm}\,&  Figure 2: The
Potential, $a$\\
\end{tabular}

\vspace{7 mm}

At about one half the Compton wavelength$ ^\dagger$ from the center there is a
 shielding effect and the
 Coulomb nature of the central charge becomes apparent. At large distances from
the center the electrostatic charges ``cancel'' each other.
 We can calculate the magnitude of the total charge
 due to the Dirac field (see \S 3.1), with $f(0)$ calculated numerically
($f(\infty)=0$, in this case),
   \begin{eqnarray*}
e\int_{\Rnos^3} j^\alpha dS_{\alpha} &=& \frac{1}{2 e}(f(\infty)-f(0))\\
                                     &\approx &\frac{1}{2 e}11.407391
   \end{eqnarray*}
This calculation results in a charge of the
 same magnitude as the central charge.

The object is highly compact, with a radius of about a (reduced) Compton
wavelength -- see figures 4 and 6. It has an onion like structure consisting
of an infinite series of spherical shells -- the local maxima of $Q$ occur at
points, $\rho=\rho_{m}$, where $ \sin (\chi(\rho_m)) =0$, however from lemma 2
we see that $\chi$ must diverge as $\rho \to 0$ (in the present case) so there
will be an infinite number of shells.

\vspace{5 mm}

\psfig{figure=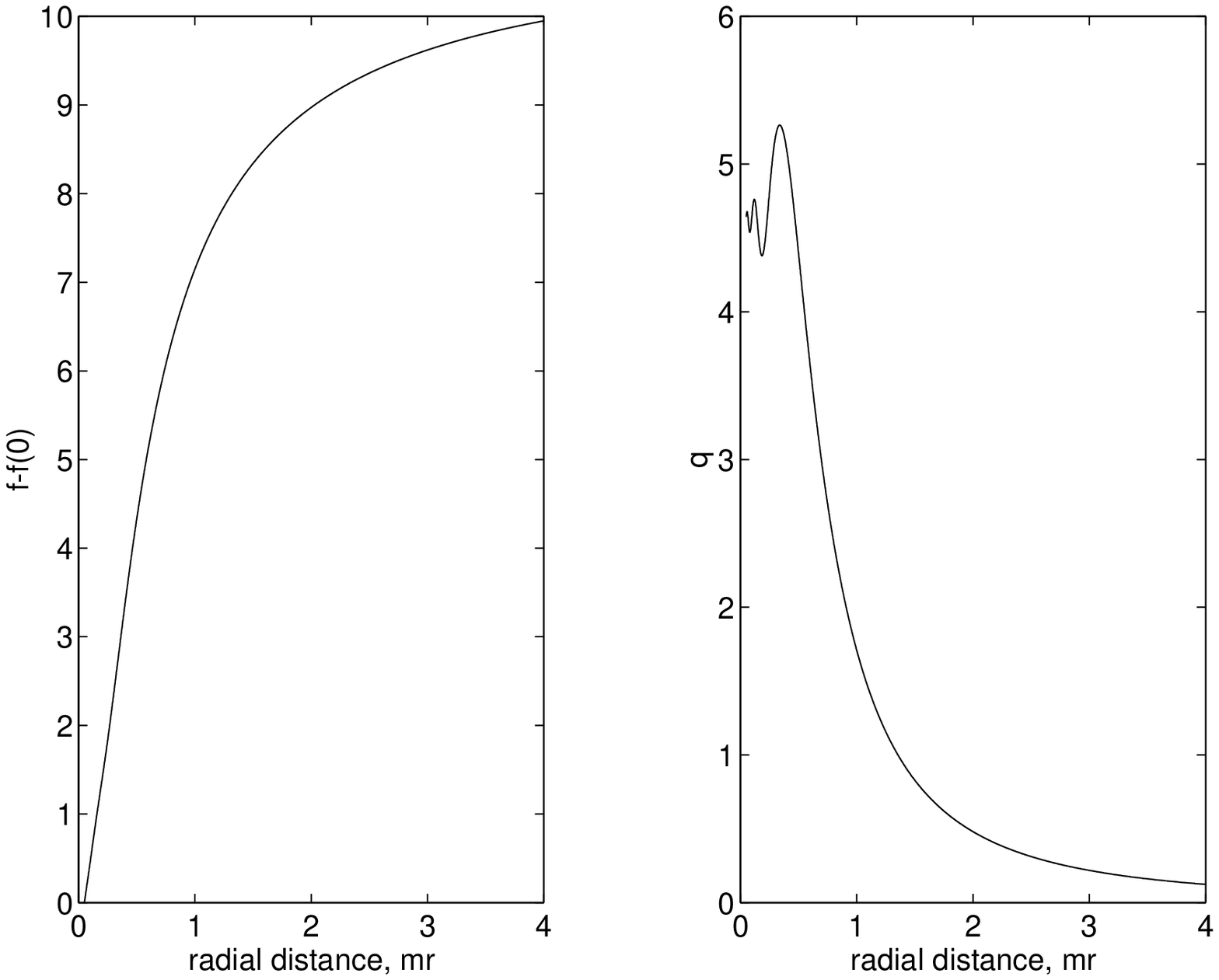,height=7cm,width=15cm}
\begin{tabular}{lll}
Figure 3: The ``Charge'' Interior&\,\hspace{1.5 cm}\,& Figure 4: The ``Charge''
on a\\
\hspace{1.8 cm}to $B(r)$, $f-f(0)$.&\,\hspace{1.5 cm}\,& \hspace{1.8 cm}Shell,
radius $mr$, $q$.
\end{tabular}

\vspace{7 mm}

We note that, in the present case, $a_{\infty}=0$ so the Dirac field
has time dependence $e^{-imt}$ -- in the gauge for which $a \to 0$ as $\rho \to
0$. The Dirac field has mass
\begin{eqnarray*}
m \int_{\Rnos ^3} j^{\alpha} dS_{\alpha}&=&\frac{m}{2 e^2}(f(\infty)-f(0)) \\
                                        &\approx&11.407391\frac{m}{2 e^2}
\end{eqnarray*}
\vspace{5 mm}

\psfig{figure=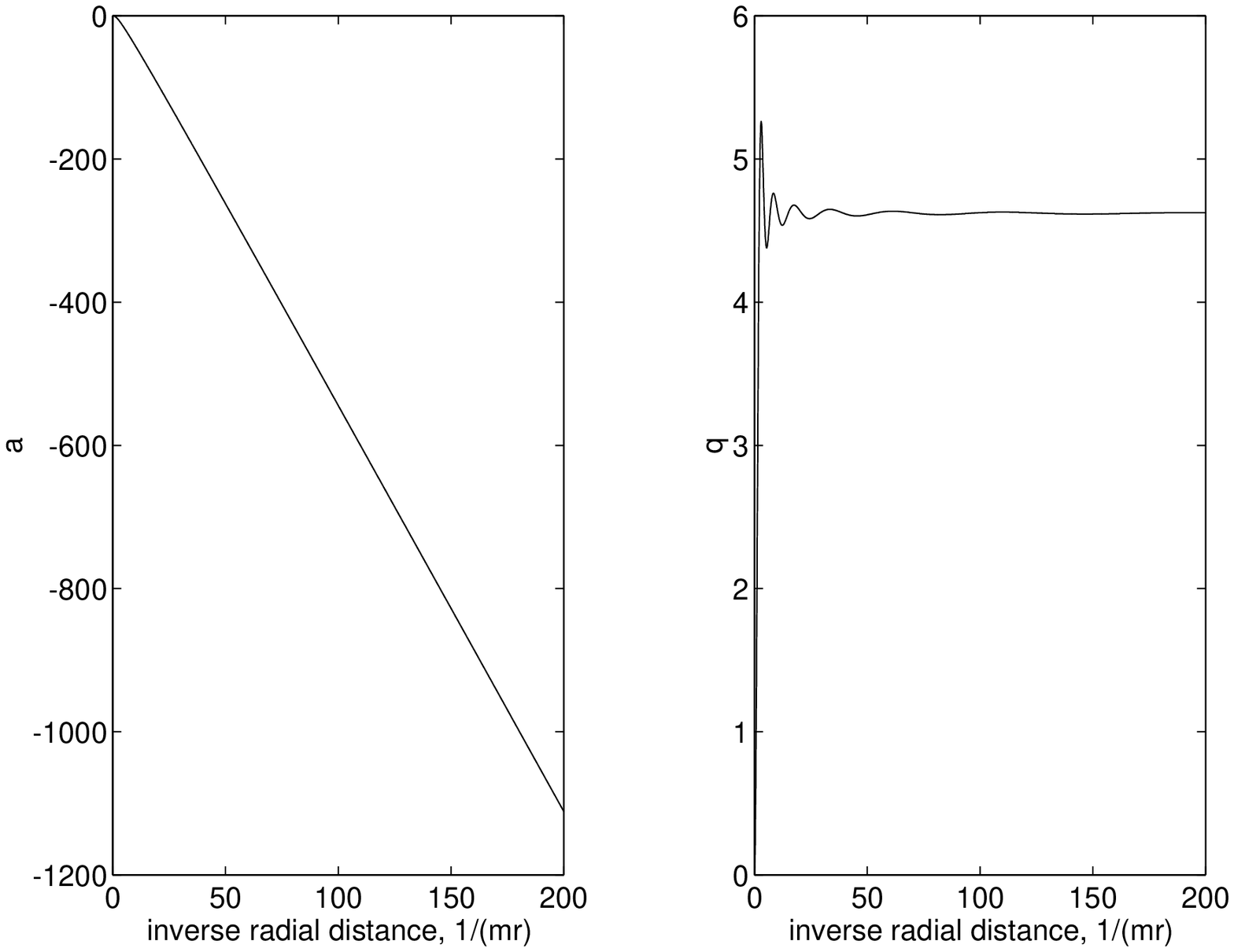,height=7cm,width=15cm}
\begin{tabular}{lll}
\hspace{.4 cm}Figure 5: The Potential, $a$,&\,\hspace{2.3 cm}\,&Figure 6:
 The Shell ``Charge'', \\
\hspace{2.1 cm} from Infinity.&\,\hspace{2.3 cm}\,& \hspace{1.8 cm}$q$, from
Infinity.
\end{tabular}

\vspace{7 mm}

To finish we calculate the energy-momentum density of the system.
 The symmetric energy-momentum tensor is
\begin{eqnarray*}
T_{\alpha \beta}&=&T^D_{\alpha \beta}+T^{em}_{\alpha \beta}\,\mbox{, with}\\
T^D_{\alpha \beta}&=&\frac{i}{4}[ \sigma_{\alpha}^{A\dot{A}}
(\overline{u}_{\dot{A}}u_{A,\beta}+v_{A}\overline{v}_{\dot{A},\beta})
+\sigma_{\beta}^{A\dot{A}}
(\overline{u}_{\dot{A}}u_{A,\alpha}+v_{A}\overline{v}_{\dot{A},\alpha})\\
\mbox{}& &-\sigma_{\alpha}^{A\dot{A}}
(u_{A}\overline{u}_{\dot{A},\beta}+\overline{v}_{\dot{A}}v_{A,\beta})
-\sigma_{\beta}^{A\dot{A}}
(u_{A}\overline{u}_{\dot{A},\alpha}+\overline{v}_{\dot{A}}v_{A,\alpha})]
+eA_{(\alpha}j_{\beta)}\\
T^{em}_{\alpha \beta}&=&-\frac{1}{4\pi}\left( F_{\alpha \gamma}
F_{\beta}^{\,\gamma}-\frac{1}{4}\eta_{\alpha \beta}F_{\mu \nu}F^{\mu
\nu}\right)
\end{eqnarray*}
These expressions are derived from the Lagrangian
\begin{eqnarray*}
\lefteqn{L=\frac{i}{2}\left(
\overline{u}_{\dot{A}}\p ^{A\dot{A}} u_{A}-u_{A}\p ^{A\dot{A}}
\overline{u}_{\dot{A}}
-\overline{v}_{\dot{A}}\p ^{A\dot{A}} v_{A}+v_{A}\p ^{A\dot{A}}
\overline{v}_{\dot{A}}\right)}\\
&&-\frac{m}{\sqrt{2}}\left( u_{A}v^{A}+\overline{u}_{\dot{A}}
\overline{v}^{\dot{A}}
\right)+ej_{\alpha}A^{\alpha}-\frac{1}{16\pi}F_{\alpha \beta}F^{\alpha \beta}
\end{eqnarray*}
Notice that in the absence of the electromagnetic field,
 for a Dirac field with time dependence
$e^{-iEt}$, the energy density $T^D_{00}$ is
          \[T^D_{0 0}=Ej_{0}\]
In the present case we have
\begin{eqnarray*}
T^D_{00}&=&j_{0}(m+eA_{0})=\frac{m^2}{e^2}\frac{Q}{4\pi r^2}(1+a)\\
T^{em}_{00}&=&\frac{m^2}{4 \pi r^2 e^2}\left\{ \left[\frac{\rho ^2}{2}
 \left(\frac{da}{d\rho}\right)^2+\frac{1}{4\rho ^2}\right]
 + \frac{\rho^2}{2}\left( \frac{da}{d\rho}\right) ^2
+\frac{1}{4\rho^2} \right\}
\end{eqnarray*}
These expressions include terms due to the central Coloumb and magnetic
monopole
fields, they lead to an infinite total energy when integrated over $\Rnos ^3$,
 the energy can, however, be regularised by removing these singular terms.
 If we exclude the energy due to the interaction between the electromagnetic
and Dirac fields, $eA_{(\alpha}j_{\beta)}$, then $T^D_{00}$ gives the mass
density $mj_{0}$ as above -- however, this procedure is clearly not gauge
invariant.

Finally, it is perhaps worth mentioning that the highly localised ``multi-
electron fields" described here may in fact have applications to objects
described in recent experimental work \cite{e:ball}
 -- ``geonium" or ``kilo-e" objects --
consisting of highly localised (point like, from the experimental viewpoint)
collections of electrons in atomic traps.

\newpage

$ ^\dagger$ We assume that the constants appearing in the Dirac equation, i.e.
$e$ and $m$ have there usual meaning -- $e$ the square root of the fine
structure constant and $m$ the inverse of the (reduced) Compton wavelength.

\noindent {\em Acknowledgements}. The author acknowledges the help and advice
of
 Gary Bunting on suitable numerical ode integrators. The author would also
 like to acknowledge Robert Bartnik for helpful discussions and advice.

\end{document}